\documentclass[notitlepage,a4paper,prd,noshowpacs,noshowkeys,nofootinbib]{revtex4}
\usepackage[latin1]{inputenc}
\usepackage{ae,aecompl}
\usepackage{amsmath,amssymb}
\usepackage{mathrsfs}
\usepackage{graphicx}
\usepackage[usenames,dvipsnames]{color} 
\usepackage{hyperref}
\hypersetup{
   colorlinks=true,       
    linkcolor=Blue,          
    citecolor=Plum,        
    filecolor=magenta,      
    urlcolor=YellowOrange           
}

\usepackage{etoolbox}
\usepackage{bookmark}
\pretocmd{\section}{\bookmarksetup{startatroot} }{}{}%

\providecommand{\hx}{\hat{x}}
\providecommand{\bx}{\mathbf{x}}
\providecommand{\bp}{\mathbf{p}}
\providecommand{\bq}{\mathbf{q}}
\providecommand{\bu}{\mathbf{u}}
\providecommand{\br}{\mathbf{r}}
\providecommand{\tb}{{\mathrm{TB}}}

\providecommand{\cp}{\mathsf{CP}}
\providecommand{\bM}{\bar{M}}

\providecommand{\om}{\omega}
\providecommand{\tS}{\tilde{S}}

\providecommand{\mns}{\mathrm{MNS}}

\providecommand{\bbs}[1]{\bar{\bs{#1}}}

\providecommand{\FF}{\mathbb{F}}
\providecommand{\mss}[1]{\mbox{\scriptsize $#1$}}
\providecommand{\ml}[1]{\mbox{\large $#1$}}
\providecommand{\tp}{{\mss{\mathsf{T}}}}
\providecommand{\ZZ}{\mathbb{Z}}
\providecommand{\ponto}{{\cdot}}
\providecommand{\su}[1]{{SU(#1)}}
\providecommand{\aver}[1]{\langle #1 \rangle}
\providecommand{\eq}[1]{\begin{equation} #1 \end{equation}}
\providecommand{\eqali}[1]{\begin{equation}\begin{aligned} #1
    \end{aligned}\end{equation}}
\providecommand{\mtrx}[1]{\begin{pmatrix} #1 \end{pmatrix}}
\providecommand{\bs}[1]{\boldsymbol{#1}}
\providecommand{\ums}[2][1]{\ml{\tfrac{#1}{#2}}} 
\providecommand{\xlink}[1]
  {\href{http://arxiv.org/abs/#1}{arXiv:#1}}

\DeclareMathOperator{\aut}{Aut}
\DeclareMathOperator{\out}{Out}

\DeclareMathOperator{\diag}{\mathrm{diag}} 
\usepackage{bbm}
\providecommand{\id}{{\mathbbm{1}}} 

\begin{document}
\title{
Generalized CP symmetries in $\Delta(27)$ flavor models
}
\author{C.~C.~Nishi}
\email{celso.nishi@ufabc.edu.br}
\affiliation{
Universidade Federal do ABC - UFABC, \\
09.210-170,
Santo André, SP, Brazil
}
\begin{abstract}

We classify explicitly all the possible generalized CP symmetries that are
definable in $\Delta(27)$ flavor models. In total, only 12 transformations are
possible. We also show interesting consequences of considering some of them as
residual symmetries of the neutrino sector.

\end{abstract}
\maketitle
\section{Introduction}
\label{sec:intro}

Neutrino physics has entered a new era after the discovery of nonzero and
relatively large $\theta_{13}$ mixing angle\,\cite{dayabay}.
Such a discovery has enabled us to pursue the determination of yet another unknown
quantity of the standard three-family description of lepton flavor physics: the
Dirac CP phase $\delta_D$. 
For Majorana neutrinos, in addition to this phase, three unknowns remain for
the complete description of the Pontecorvo-Maki-Nakagawa-Sataka (PMNS) mixing 
matrix: the two Majorana CP phases and the neutrino mass hierarchy.
The determination of $\delta_D$ might be possible in the foreseeable future, and 
hints of a nonzero $\delta_D$ are starting to show up in global fits of oscillation
parameters\,\cite{fogli.12,valle.12,GG:12}.

On the theoretical side, a relatively large $\theta_{13}$ angle discards the exact
validity of certain mass-independent textures\,\cite{lam:texture} for the PMNS
matrix, the most popular being the tribimaximal (TBM) form\,\cite{tbm}, which was a
good approximation until recently.
The great appeal of these mass-independent textures comes from the fact that
they can arise naturally as a consequence of non-abelian discrete flavor symmetries
that act on the horizontal space of the three families of leptons.
The general scheme consists of assuming a non-abelian discrete flavor group $G_F$,
which is broken into different subgroups on the charged lepton and neutrino sectors.

If we believe that discrete flavor symmetries govern the observed pattern of
the mixing angles and mass hierarchies of leptons, including the observed value of
the $\theta_{13}$ angle, we need to modify the simple forms, such as the TBM form,
by either (i) adding corrections or (ii) considering different symmetries. 
One route is to consider the minimal amount of residual symmetries on the mass
matrices which still remains predictive\,\cite{hernandez,dicus.12,hu}.
Here we pursue (ii) by employing generalized CP (GCP)
transformations\,\cite{grimus:gcp}.
We consider the possibility of reducing the symmetry of the neutrino sector
to a \textit{single} GCP symmetry. We will see that such a setting leaves a lot of
freedom in the leptonic mixing matrix but it is compatible with a more general
scenario where some other horizontal symmetry is \textit{approximately} valid in the
neutrino sector instead of being exactly satisfied at leading order.
Obviously, it is possible to account for nonzero $\theta_{13}$, even in the symmetry
limit, if we consider more complicated flavor groups\,\cite{toorop}.

One successful modification of the residual symmetry of the neutrino sector consists
of replacing the familiar $\mu\tau$ interchange symmetry\,\cite{mutau} (in the
flavor basis) with the symmetry called $\mu\tau$ reflection\,\cite{mutau:r}, which
corresponds to the joint application of $\mu\tau$ interchange together with
complex conjugation of $M_\nu$. This symmetry leads to maximal $\theta_{23}$ and
maximal Dirac CP phase by allowing nonzero but free
$\theta_{13}$\,\cite{grimus:gcpnu}. This symmetry has been successfully implemented
in a number of models\,\cite{mutau:r:models,s4tilde}.

The possibility of considering generalized CP transformations\,\cite{grimus:gcp} as
symmetries in the leptonic sector has been
analyzed recently\,\cite{hagedorn:gcp,lindner:gcp}. The work of
Ref.\,\cite{hagedorn:gcp} focuses on analyzing the consequences of having a residual
GCP symmetry in the neutrino sector along with other residual horizontal symmetries
in the charged lepton and neutrino sectors. In particular, the flavor groups $S_4$
and $A_4$ have been analyzed and implemented recently\,\cite{hagedorn:13}.

On the other hand, the role of GCP symmetries as automorphisms of the
(horizontal) flavor group is studied in Ref.\,\cite{lindner:gcp}.
The authors develop the general theory and then analyze the relevant cases
from the literature. 
For example, they show that there is only one possible nontrivial definition for GCP
within $A_4$ models which leads to the $\tS_4$ flavor symmetry of
Ref.\,\cite{s4tilde}.

In this work, we further consider the possible GCP symmetries that are definable in
$\Delta(27)$ flavor models.\footnote{The flavor group $\Delta(27)$ was first
considered for quarks in Ref.\,\cite{geometric.CP} and for leptons in
Ref.\,\cite{D27:models}.}
Such a flavor group is interesting from the point of view of GCP symmetries because
it possesses a large number of automorphisms that can be used as GCP
symmetries\,\cite{lindner:gcp}.
At the same time, the group does not possess any order-2 element that could be
promoted to a residual symmetry of the neutrino mass matrix.
The generators of the automorphism group for $\Delta(27)$ were given in
Ref.\,\cite{lindner:gcp}, but we intend here to find all possible GCP symmetries
explicitly and consider constraints that were not discussed previously.
Furthermore, we analyze the possibility of considering these GCP symmetries as
residual symmetries of the neutrino sector.

The outline of this work is the following: In Sec.\,\ref{sec:residual}, we review
the consequences of having only a single GCP symmetry as a residual symmetry of the
neutrino mass matrix. We list all the possible GCP symmetries in $\Delta(27)$
flavor models in Sec.\,\ref{sec:list} and extract some interesting features.
Section \ref{sec:gcp} reviews the consequences of adding one GCP symmetry in a
theory invariant by a discrete flavor symmetry.
We justify the list of possible GCP symmetries of $\Delta(27)$ models in
Sec.\,\ref{sec:D27}.
The conclusions are shown in Sec.\,\ref{sec:conclusion}.

\section{Residual GCP symmetries}
\label{sec:residual}

Let us start our study of the consequences of residual GCP symmetries on the
mass matrices by first reviewing here the consequences of the usual (unitary)
residual symmetries  acting on the neutrino mass matrix as\,\cite{lam:s4}
\eq{
\label{S:hor}
S^\tp M_\nu S=M_\nu\,.
}
We assume the charged lepton mass matrix squared $\bM_l\equiv M_lM_l^\dag$ is
diagonal (flavor basis), which should be ensured by another residual symmetry
$G_l$.

We first recall that if $U$ is the matrix that diagonalizes $M_\nu$ as
\eq{
\label{diag:Mnu}
U^\tp M_\nu U=\diag(m_i),~~m_i>0,
}
with nonzero and nondegenerate masses $m_i$,
then any other matrix $U'$ that also diagonalizes $M_\nu$ must be related to $U$
by\,\cite{lam:s4,grimus:s4}
\eq{
U'=Ud\,,
}
where $d$ is a diagonal matrix with nonzero entries $\pm 1$.

The symmetry \eqref{S:hor} dictates that if $U$ diagonalizes $M_\nu$, then $SU$ also
diagonalizes it, and then
\eq{
\label{SU:hor}
SU=U d\,.
}
This means that the eigenvectors of $M_\nu$ are also eigenvectors of $S$ with
eigenvalues $\pm 1$. Furthermore, the symmetry matrix $S$ fixes one eigenvector
of $M_\nu$ corresponding to the unique nondegenerate eigenvalue of $S$;  we
obviously exclude the cases $S=\id$ or $S=-\id$.
On the other hand, the property $d^2=\id$ implies $S^2=\id$, and only $\ZZ_2$
symmetries can be implemented on $M_\nu$, the maximal symmetry being
$\ZZ_2\times\ZZ_2$, generated by two matrices $S_1,S_2$.

Instead of considering the unitary symmetry \eqref{S:hor}, we can consider the
antiunitary symmetry:
\eq{
\label{S:gcp}
S^\tp M_\nu S=M_\nu^*\,.
}
For example, the choice
\eq{
S=\mtrx{1&0&0\cr0&0&1\cr0&1&0}\,,
}
corresponds to $\mu\tau$-reflection symmetry\,\cite{mutau:r,grimus:gcpnu}.

The symmetry \eqref{S:gcp} now implies that $SU^*$ also diagonalizes $M_\nu$ if $U$
does, and then \eqref{SU:hor} is replaced by
\eq{
\label{SU:gcp}
SU^*=Ud_\nu\,.
}
This is equivalent to saying that $S=Ud_\nu U^\tp$, and then necessarily
\eq{
\label{S:T}
S^\tp =S\,,
}
if we require nonzero and nondegenerate neutrino masses.
Since $S$ is also unitary, it obeys $S^*S=SS^*=\id$.
Note that the symmetry condition \eqref{S:T} is invariant by basis change.

We can show that any unitary and symmetric matrix $S$ can be diagonalized by a real
orthogonal matrix $R$ as
\eq{
\label{S:diag}
R^\tp S R=\eta\equiv \diag(\eta_i)\,,~~|\eta_i|=1\,.
}
Another way of writing \eqref{S:diag} is
\eq{
SR=R\eta\,.
}
The proof consists of writing $S=S_1+iS_2$, where $S_1,S_2$ are real symmetric
matrices.
Then $S^*S=\id$ implies that $S_1$ commutes with $S_2$ and they can be
simultaneously diagonalized by $R$.

Now let us denote by $\bu_i$ the columns of $U$.
The relation \eqref{SU:gcp} implies
\eq{
\label{Sui}
S\bu_i^*=\pm \bu_i\,,
}
the sign $\pm 1$ being given by $(d_\nu)_{ii}$.
We can choose all $\bu_i$ to obey the plus-sign equation of \eqref{Sui} by
conveniently replacing $\bu_i$ with $i\bu_i$ when $(d_\nu)_{ii}=-1$. This leads to
\eq{
\label{U:1}
U=\left(\begin{array}{c|c|c}\bu_1 & \bu_2 & \bu_3
\end{array}\right)d_\nu^{\,\frac{1}{2}}\,;
}
we choose $(d_\nu^{\,\frac{1}{2}})_{ii}=i$ if
$(d_\nu)_{ii}=-1$ or $(d_\nu^{\,\frac{1}{2}})_{ii}=1$ if $(d_\nu)_{ii}=1$.

We now expand $\bu_i$ in terms of the real eigenvectors $\br_i$ of $S$,
corresponding to the columns of $R$,
\eq{
\bu_i=\br_ja^j_i\,.
}
Equation \eqref{Sui} with the plus sign leads to
\eq{
\eta_j(a^j_i)^*=a^j_i\,.
}
This is solved by
\eq{
\bu_i=\br_i\eta_j^{\frac{1}{2}}a^j_i\,,~~\text{with real $a^j_i$},
}
where we have made the replacement $a^j_i\to \eta_j^{\frac{1}{2}}a^j_i$.
Equation \eqref{U:1} is finally
\eq{
U=R\,\eta^{\frac{1}{2}} O_\nu\,d_\nu^{\frac{1}{2}}\,,
}
where $O_\nu$ is a real orthogonal matrix defined by $(O_{\nu})_{ji}=a^j_i$.
Orthogonality of $O_\nu$ follows from unitarity of $U$.
The combination
\eq{
\label{U_S}
U_S=R\,\eta^{\frac{1}{2}}
}
is uniquely determined by $S$ (except for sign ambiguities) if $S$ is
nondegenerate.

In the flavor basis, $V_{\mns}=U_\nu$ and then the lepton mixing matrix,
\eq{
\label{MNS}
V_{\mns}=R\,\eta^{\frac{1}{2}} O_\nu\,d_\nu^{\frac{1}{2}}\,,
}
is determined by the antiunitary symmetry \eqref{S:gcp}, if $S$ is nondegenerate, up
to the three-parameter freedom of choosing $O_\nu$.
In particular, the CP properties are completely determined by $\eta^{\frac{1}{2}}$
and $d_\nu^{\frac{1}{2}}$.
Therefore, among the 6 parameters of the PMNS matrix, the 3 phases are
indirectly determined by the symmetry.
The presence of another additional unitary symmetry as \eqref{S:hor}, that commutes
with the antiunitary symmetry \eqref{S:gcp}, fixes one column of the matrix
$O_\nu$\,\cite{hagedorn:gcp}.
However, compared to Ref.\,\cite{hagedorn:gcp}, the form in Eq.\,\eqref{MNS} shows
more explicitly the separate dependence of the PMNS matrix on the fixed phases
($\eta^{\frac{1}{2}}$) and real elements ($R$). 
In a general basis, the form \eqref{MNS} needs to be adapted to show explicit
dependence on the residual symmetry of the charged lepton sector.

The desired setting is the following: if somehow $R$ can be chosen close to the
experimental mixing matrix, then $O_\nu$ can be close to the identity and treated as
a perturbation.

\section{Possible GCP symmetries in $\Delta(27)$ models}
\label{sec:list}

We seek now some possible GCP symmetries which could be phenomenologically
interesting. We choose $\Delta(27)$ as the flavor group because it possesses a large
amount of possible nontrivial GCP symmetries\,\cite{lindner:gcp}.
See Ref.\,\cite{D27:models} for the first applications of the $\Delta(27)$ flavor
group in the lepton sector.

The group $\Delta(27)\simeq(\ZZ_3\times\ZZ_3)\rtimes\ZZ_3$ is an order-27
non-abelian
finite group which can be defined by using two generators $a,b$ and another
auxiliary element $a'$ through the relations\,\cite{ishimori}
\eq{
\label{D27:present}
\begin{gathered}
a^3={a'}^3=b^3=e,~~aa'=a'a\,,\cr
bab^{-1}=(aa')^{-1},~~ba'b^{-1}=a\,.
\end{gathered}
}
Note that $a,a'$ generate the invariant subgroup $\ZZ_3\times\ZZ_3$ and the
element,
\eq{
\label{z0}
z_0\equiv a{a'}^{-1}\,,
}
generates the center of the group $\mathcal{Z}(\Delta(27))\simeq\ZZ_3$.

In three dimensions we can use the explicit (faithful) representation $\bs{3}$ for
$\Delta(27)$:
\eq{
\label{D27:3}
D_3(b)=T\equiv\mtrx{0&0&1\cr 1&0&0\cr 0&1&0},~~
D_3(a)=\diag(1,\om,\om^2),~~
D_3(a')=\diag(\om,\om^2,1).
}
This representation differs slightly from that in Ref.\,\cite{ishimori}.
Notice that $Te_i=e_{\sigma(i)}$ where $\sigma=(123)$. This means that, on a vector
$x=(x_1,x_2,x_3)^\tp$, $x\to Tx$ induces $x_i\to x_{\sigma^{-1}(i)}$, i.e., the
permutation $(132)$.

The definable GCP symmetries for any flavor group were studied in
Ref.\,\cite{lindner:gcp} as automorphisms acting on the flavor group.
Although the generators of the automorphism group were listed there, the
possible GCP symmetries were not listed explicitly.
Here we show in Sec.\,\ref{sec:D27} that there are only 12 possible nonequivalent
GCP symmetries that can be defined within a $\Delta(27)$ flavor group for the tree
family of left-handed leptons $L_i$ transforming as $\bs{3}$ in \eqref{D27:3}.
They are given by
\eq{
\label{gcp:L}
L_i\to (S_k)_{ij}(CL_j^*)\,,
}
where
\eqali{
\label{list:S}
S_0&=\id_3\,,& S_1&=\mtrx{1&0&0\cr0&0&1\cr0&1&0}\,,&
  S_2&=\diag(1,1,\om)\,,&S_3&=\diag(1,1,\om^2)=S_2^2\,,
\cr
S_4&=U_\om\,,& S_5&=U_\om^*=S_4S_1\,,&
  S_6&=\frac{-i\om}{\sqrt{3}}\mtrx{\om^2&1&1\cr1&1&\om^2\cr1&\om^2&1}\,,&
  S_7&=\frac{i\om^2}{\sqrt{3}}\mtrx{\om&1&1\cr1&1&\om\cr1&\om&1}\,,
\cr
S_8&=\mtrx{\om&0&0\cr0&0&1\cr0&1&0}\,,&
  S_9&=\mtrx{\om^2&0&0\cr0&0&1\cr0&1&0}\,,& 
  S_{10}&=S_6S_1\,,& S_{11}&=S_7S_1\,.
}

The GCP symmetries \eqref{list:S} are unique up to composition with elements of
$\Delta(27)$ itself and multiplication by an overall phase factor.
Concerning the first freedom, we choose $S_i$ to be symmetric\,\footnote{In
$\Delta(27)$, this choice is always possible; see Sec.\,\ref{sec:obtaining}.}
\eqref{S:T} so that they can be used as residual symmetries of $M_\nu$.
We consider only the GCP symmetries that do not enlarge the horizontal flavor
group $\Delta(27)$; see discussion in Sec\,.\ref{subsec:gcp^2}.
The transformation properties for the singlets $\bs{1}_{rs}$ are fixed according
to the automorphism these matrices induce on $\Delta(27)$\,\cite{lindner:gcp}.
Sometimes one singlet cannot appear alone but has to be paired up with another
singlet.

In principle, we can use all $S_i$ with $i=0,\ldots,11$ as residual GCP symmetries
for $M_\nu$.
However, $S_0$ corresponds to the usual CP transformation and is therefore
noninteresting.
The GCP symmetry corresponding to $S_1$ is interesting, but it
corresponds to $\mu\tau$ reflection, which was considered previously in the
literature, e.g., in Ref.\,\cite{s4tilde}. The matrices $S_2,S_3$ are diagonal, so
they have trivial eigenvectors. Analogously, $S_8,S_9$ are block diagonal, so they
have one trivial eigenvector.
The matrices $S_{10},S_{11}$ should also be discarded because they have
degenerate eigenvalues\,\footnote{The unitary version of $S_{11}$ was used in a
different context in Ref.\,\cite{branco:demo}.}. 
The remaining $S_4,S_5,S_6,S_7$ are potential candidates for further study.

The GCP transformations corresponding to $S_4,S_6,S_5,S_7$ are potentially
interesting, because the matrix \eqref{U_S} are given by
\eqali{
\label{US}
U_{S_4}&=\left(
\begin{array}{ccc}
 \frac{1+\sqrt{3}}{\sqrt{2 \left(3+\sqrt{3}\right)}} & \frac{
 \left(1-\sqrt{3}\right)}{\sqrt{6-2 \sqrt{3}}} & 0 \\
 \frac{1}{\sqrt{2 \left(3+\sqrt{3}\right)}} & \frac{1}{\sqrt{6-2 \sqrt{3}}} &
-\frac{1}{\sqrt{2}} \\
 \frac{1}{\sqrt{2 \left(3+\sqrt{3}\right)}} & \frac{1}{\sqrt{6-2 \sqrt{3}}} &
\frac{1}{\sqrt{2}}
\end{array}
\right)
\diag(1,i,e^{i\pi/4})
\,,&
U_{S_6}&=U_\tb\diag(e^{i2\pi/3},1,e^{i\pi/6})\,.
\\
U_{S_5}&=\left(
\begin{array}{ccc}
 \frac{1+\sqrt{3}}{\sqrt{2 \left(3+\sqrt{3}\right)}} & \frac{
 \left(1-\sqrt{3}\right)}{\sqrt{6-2 \sqrt{3}}} & 0 \\
 \frac{1}{\sqrt{2 \left(3+\sqrt{3}\right)}} & \frac{1}{\sqrt{6-2 \sqrt{3}}} &
-\frac{1}{\sqrt{2}} \\
 \frac{1}{\sqrt{2 \left(3+\sqrt{3}\right)}} & \frac{1}{\sqrt{6-2 \sqrt{3}}} &
\frac{1}{\sqrt{2}}
\end{array}
\right)
\diag(1,i,e^{-i\pi/4})
\,,&
U_{S_7}&=U_\tb\diag(e^{-i2\pi/3},1,e^{-i\pi/6})\,,
}
where $U_{\tb}$ is the familiar tribimaximal mixing matrix,
\eq{
\label{def:UTB}
U_\tb=
\begin{pmatrix}
\sqrt{\frac{2}{3}} & \ums{\sqrt{3}} & 0\cr
-\ums{\sqrt{6}} & \ums{\sqrt{3}} & -\ums{\sqrt{2}}\cr
-\ums{\sqrt{6}} & \ums{\sqrt{3}} & \ums{\sqrt{2}}
\end{pmatrix}\,.
}
We have chosen the order of the eigenvectors appropriately.
Note that the tribimaximal matrix appears for $S_6$ and $S_7$.
Numerically, however, all matrices are close, as
\eq{
|U_{S_4}|=|U_{S_5}|=\left(
  \begin{array}{ccc}
  0.888074 & 0.459701 & 0 \\
  0.325058 & 0.627963 & 0.707107 \\
  0.325058 & 0.627963 & 0.707107
  \end{array}
  \right)
\,,\quad
|U_{S_6}|=|U_{S_7}|=\left(
  \begin{array}{ccc}
  0.816497 & 0.57735 & 0 \\
  0.408248 & 0.57735 & 0.707107 \\
  0.408248 & 0.57735 & 0.707107
  \end{array}
  \right)\,.
}

Instead of $S_{10}$ and $S_{11}$, which we discarded because of degenerate
eigenvalues, we could have considered $D_3(a)S_{10}D_3(a)$ and $D_3(a)S_{11}D_3(a)$
or their composition with $D_3(a^2)$.
They have nondegenerate eigenvalues and are symmetric, but they lead to either
\eq{
|U_S|=
\left(
\begin{array}{ccc}
 0.84403 & 0.449099 & 0.293128 \\
 0.293128 & 0.84403 & 0.449099 \\
 0.449099 & 0.293128 & 0.84403
\end{array}
\right)
\text{ or }
\left(
\begin{array}{ccc}
 0.84403 & 0.449099 & 0.293128 \\
 0.449099 & 0.293128 & 0.84403 \\
 0.293128 & 0.84403 & 0.449099
\end{array}
\right)\,,
}
which inevitably lead to a large $\theta_{13}$ angle.
This case reminds us that the composition with elements of the horizontal symmetry
do lead to different physical predictions if GCP transformations are considered as
residual symmetries\,\cite{hagedorn:gcp}.

\section{Inclusion of a GCP transformation}
\label{sec:gcp}

To obtain all the possible GCP symmetries listed in \eqref{list:S} which are
consistent with the flavor group $\Delta(27)$, we need to study how to extend 
a discrete symmetry group $G_H$ by the inclusion of \textit{one} generalized CP
transformation (GCP) acting, e.g., as \eqref{gcp:L} for the three families of
left-handed leptons.
This study was performed in general in Ref.\,\cite{lindner:gcp}.
Here we consider it in more detail and additionally add more constraints not
previously considered.

\subsection{CP as automorphism}

We begin by reviewing how GCP transformations induce an automorphism on other
symmetry groups of the theory, especially on discrete
symmetries\,\cite{lindner:gcp}.

Let a discrete group, $G_H$\,\footnote{$H$ stands for horizontal.} act on the
scalar multiplet of fields $\phi$ as 
\eq{
\phi\to D(g)\phi\,,
}
where $g\in G_H$ and $D$ is a (possibly reducible) representation of $G_H$.
This setting can be easily extended to other nonscalar fields.

A generalized CP transformation (GCP) acts as
\eq{
\label{phi:gcp}
\phi\to \tS\ponto\phi\equiv S\phi^*(\hx)\,,
}
where $\hat{x}=(x_0,-\bx)$ if $x=(x_0,\bx)$.
Notice that $S$ should be unitary to preserve the kinetic term.
For fermionic fields, it is implicit that we factor $\cp^2=-1$.

Invariance of the theory by $\tS$ and $G_H$ leads to an invariance by the
composition
\eq{
\label{cp:auto}
\phi\stackrel{\tS}{\longrightarrow} S\phi^*
  \stackrel{g}{\longrightarrow}SD(g)^*\phi^*
  \stackrel{{\tS}^{-1}}{\longrightarrow}SD(g)^*S^{-1}\phi\,,
}
which is a horizontal (unitary) transformation.
The last transformation in \eqref{cp:auto} should be an element of $D(G_H)$
because otherwise we would have to enlarge $G_H$.
Hence, by defining
\eq{
\label{D_S}
D_{\tS}(g)\equiv SD(g)^*S^{-1}\,,
}
it is required that there always exist some $g'\in G_F$ such that
\eq{
\label{D_S=D}
D_{\tS}(g)=D(g'),~ \text{ for all $g\in G_F$}.
}

We can easily show that $D_S$ is also a representation for
$G_H$\,\cite{lindner:gcp}.
Moreover, $\ker D_S=\ker D$ and then $D_S$ is faithful if $D$ is faithful.
Considering that the representation $D$ is \textit{faithful}, the mapping
$\tau\equiv D^{-1}\circ D_S$ exists (restricted to the image of $D$ and $D_S$) and
is a homomorphism between $G_H$ and itself:
\eq{
g\to g'=\tau(g)\,.
}
Since $D_S$ is also faithful, $\tau$ is invertible and it is then an
\textit{automorphism} between $G_H$ and itself.
The possible matrices $S$ in \eqref{D_S} then realize some element of the
automorphism group $\aut(G_H)$.
We can then rewrite the condition \eqref{D_S=D} as
\eq{
\label{D_S=D:tau}
D_{\tS}(g)=D(\tau(g)), 
}
for all $g\in G_F$ and some automorphism $\tau$.

Suppose now there is a matrix $S=S(\tau)$ which solves \eqref{D_S=D:tau} for some
automorphism $\tau$. We can see that the matrix $S'S$, where $S'=D(g')$ corresponds
to a group element, also solves \eqref{D_S=D:tau} for the automorphism
$c_{g'}\circ\tau$ since
\eq{
D_{S'\tS}(g)=(S'S)D(g)^*(S'S)^{-1}=S'D(\tau(g))S'^{-1}
  =D(g'\tau(g)g'^{-1})=D(c_{g'}\circ\tau(g));
}
we have defined the conjugation by the element $g'$ as
\eq{
\label{conj}
c_{g'}(g)\equiv g'gg'^{-1}\,.
}
The automorphism generated by conjugation as in Eq.\,\eqref{conj} is denoted as
\textit{inner} whereas the automorphism that is not inner is called
\textit{outer}. All the inner automorphisms compose the inner autormophism group
$\mathrm{Inn}(G_H)$, an invariant subgroup of $\aut(G_H)$.
Given that conjugation by group elements trivially corresponds to an automorphism,
we only need to consider the outer automorphism group defined by
\eq{
\label{outer}
\out(G_H)\equiv \aut(G_H)/\mathrm{Inn}(G_H)\,.
}
At the Lagrangian level, inner automorphisms do not introduce any restriction when
we extend $G_H$ to $G_H\rtimes\aver{\mathrm{GCP}}$.

Suppose now that there are two matrices $S_0$ and $S=S_1S_0$ which satisfy
\eqref{D_S=D:tau} for a common automorphism $\tau$.
The relation between $S$ and $S_0$ is
\eq{
\label{U1}
S_1=\bigoplus_\alpha(s_\alpha\otimes \id_{d_\alpha}),
}
where $s_\alpha$ is an $m_\alpha\times m_\alpha$ unitary matrix acting on the
\textit{horizontal space} of $m_\alpha$ copies of the irreducible representation
(irrep) $\alpha$.
We are using the decomposition
\eq{
D(g)=\bigoplus_\alpha \big(\id_{m_\alpha}\otimes D_{\alpha}(g)\big)\,.
}
The proof of \eqref{U1} follows from the Schur lemma and is analogous to
Theorem 1 of Ref.\,\cite{grimus.rebelo}.
Therefore, two matrices that satisfy \eqref{D_S=D:tau} for the same automorphism
$\tau$ differ only by unitary change of basis on the horizontal space of replicated
irreps of $G_H$.

Let us analyze the case of trivial automorphism, i.e., $\tau=\mathrm{id}$.
Equation \eqref{D_S=D:tau} implies
\eq{
\label{tau=1}
SD(g)^*S^{-1}=D(g)\,,~~\text{for all $g\in G_F$}.
}
If $D$ is an irreducible representation, we can distinguish three cases:
real ($D^{(r)}$), pseudoreal ($D^{(p)}$) or complex ($D^{(c)}$) representation.
For real $D^{(r)}$, Eq.\,\eqref{tau=1} can be satisfied with $S=\id$.
For pseudoreal $D^{(p)}$, by definition, there is also a unitary (antisymmetric)
$W$ such that
\eq{
\label{pseudoreal}
WD^{(p)}(g)^*W^{-1}=D^{(p)}(g),~~\text{for all $g\in G_F$.}
}
For a single complex $D^{(c)}$, it is not possible to satisfy
\eqref{tau=1}, because $D^{(c)}$ and $D^{(c)*}$ are inequivalent.
However, for a reducible representation $D^{(c)}\oplus D^{(c)*}$, Eq.\,\eqref{tau=1}
can be satisfied as
\eq{
\label{DD^*}
\mtrx{0 & \id \cr \id & 0}
\mtrx{D^{(c)}(g) & 0 \cr 0 & D^{(c)}(g)^*}^*
\mtrx{0 & \id \cr \id & 0}
=
\mtrx{D^{(c)}(g) & 0 \cr 0 & D^{(c)}(g)^*}\,.
}
Therefore, in a $G_H$-invariant theory, invariance by the CP transformation
associated with the identity automorphism  demands the presence of a multiplet
$\psi'$ transforming as $D^{(c)*}$ if the theory contains a multiplet $\psi$
transforming as $D^{(c)}$. 
We have to keep in mind that a GCP transformation also induces an automorphism on
other groups involved such as gauge groups or the Lorentz group. 
Thus $\psi'$ should have the same quantum numbers of $\psi$ with respect to
these other groups because the GCP transformation that
leads to \eqref{DD^*} is
\eq{
\psi(x)\stackrel{GCP}{\longrightarrow} \psi^{\prime *}(\hx)\,,~~
\psi'(x)\stackrel{GCP}{\longrightarrow} \psi^{*}(\hx)\,.
}

In contrast, for gauge groups, the automorphism that customarily meets the
expectation of reversing gauge quantum numbers is the contragradient
automorphism $\psi^\Delta$\,\cite{grimus.rebelo} which can be defined for the
fundamental representation of $\su{n}$ by $S=\id$ and
\eq{
D^{(c)*}(g)=D^{(c)}(\psi^\Delta(g))\,.
}
We note that $\psi^\Delta$ is outer for $\su{n}$, $n\ge 3$, and $U(1)$.
By associating CP with $\psi^\Delta$, any gauge theory with scalars or fermions
interacting only by gauge interactions is always CP invariant\,\cite{grimus.rebelo}.

Assume now that $\tau$ has finite order\footnote{A finite group $G_H$ always
has a finite $\aut(G_H)$.} $m$, i.e., $\tau^m=\mathrm{id}$. 
Let us study the composition of \eqref{D_S=D:tau}. If we apply it twice, we
obtain
\eq{
\label{tau:2}
D(\tau^2(g))=(SS^*)D(g)(SS^*)^{-1}\,.
}
More generally, we obtain 
\eq{
\label{tau:even}
D(\tau^{2n}(g))=(SS^*)^nD(g)(SS^*)^{-n}\,,
}
if we apply it an even number of times, or
\eq{
\label{tau:odd}
D(\tau^{2n+1}(g))=\big((SS^*)^nS\big) D(g)^*\big((SS^*)^{n}S\big)^{-1}\,,
}
if we apply it an odd number of times.
We will see in Sec.\,\ref{subsec:gcp^2} that the order of the automorphism $\tau$
associated with a GCP transformation should be even.
Therefore, from the Schur lemma (and unitarity of $S$), for $m=2n$, we need 
\eq{
\label{SS^*:n:0}
(SS^*)^n=\id\,,
}
within all irrep sectors.

\subsection{Composition of GCP transformations}
\label{subsec:gcp^2}

We should analyze now the conditions imposed by the composition of the GCP
transformation $\tS$ itself.

If we apply the transformation \eqref{phi:gcp} twice, we would obtain
\eq{
\label{tS.tS}
\phi(x)\stackrel{\tS}{\longrightarrow}S\phi^*(\hx)
\stackrel{\tS}{\longrightarrow}(SS^*)\phi(x)\,.
}
This is just the statement that usual $\cp$ has order 2 (order $4$ for fermions).
However, Eq.\,\eqref{tS.tS} also implies that 
\eq{
\label{SS^*}
SS^*=D(s)\,,~~\text{for some $s\in G_H$,}
}
because otherwise $G_H$ would be larger by the symmetry represented by $SS^*$.

The requirement \eqref{SS^*} applied to \eqref{tau:2} implies
\eq{
\label{tau:2:2}
D(\tau^2(g))=D(sgs^{-1})\,,~~\text{for all $g$ in $G_H$.}
}
This means that any automorphism $\tau$ associated with a GCP transformation should
have order two, modulo inner automorphisms, i.e., $\tau^2=c_s$.
This requirement was not considered in Ref.\,\cite{lindner:gcp}.
We should emphasize that the consistency condition \eqref{SS^*} is indeed
\emph{independent} from the automorphism condition \eqref{D_S=D:tau} when 
$\tau^2=c_s$ is not automatic, i.e., when $\out(G_H)$ has elements of order greater
than two\,\footnote{The author is thankful to G.~-J.~Ding for raising this
question during FLASY2013.}. This is the case with $G_H=\Delta(27)$; see a specific
example in \eqref{tau:4:example}.
On the other hand, if $\out(G_H)$ has only elements of order at most two, the Schur
lemma applied to \eqref{tau:2} implies that the condition \eqref{SS^*} is
automatically satisfied. This is the case, e.g., of $G_H=A_4$.

Another condition coming from the finiteness of $s$ requires
\eq{
\label{SS^*:n}
(SS^*)^n=\id\,,
}
for $s^n=e$, $e$ being the identity element of $G_H$.
This relation is identical to \eqref{SS^*:n:0}.
Thus the GCP transformation $\tS$ always has even order $2n$, and it induces an
automorphism of the same even order.

Let us now rewrite \eqref{SS^*} as
\eq{
\label{tau:s}
SS^*=SD(s)^*S^{-1}=D_{\tS}(s)=D(\tau(s))\,.
}
Therefore, the automorphism $\tau$ induced by $S$ should leave the element $s$
invariant.

If we want $S$ to generate a residual GCP symmetry on the neutrino mass
matrix, $S$ needs to be symmetric \eqref{S:T} and then $SS^*=\id_3$.
Hence, \eqref{SS^*} and \eqref{tau:2:2} imply
\eq{
\label{s=e}
s=e\,,~~\tau^2=\mathrm{id}.
}
This is the case of usual CP symmetry $\tS=\cp$.

However, even if $S$ is nonsymmetric, the addition of the GCP transformation $\tS$
might be equivalent to the addition of another transformation $\tS'$ with
symmetric $S'$. Let us define
\eq{
S'\equiv D(g)S\,,
}
and calculate
\eq{
S'S^{\prime*}=D(g\tau(g)s)\,.
}
We have used \eqref{D_S=D:tau}.
Thus we obtain $S'S^{\prime*}=\id$ if we can find $g$ in $G_H$ such that
\eq{
g\tau(g)s=e\,.
}
If $s$ has odd order $2m+1$, this condition is automatically satisfied by $g=s^m$,
since
\eq{
g\tau(g)s=s^ms^ms=s^{2m+1}=e\,,
}
where \eqref{tau:s} was applied.

Additionally, we can see from \eqref{SS^*:n} that for each matrix $S$ that satisfies
\eqref{D_S=D:tau} and then corresponds to a consistent GCP transformation,
there are equally consistent choices of $S$ related by rephasing:
\eq{
\label{S:e^i}
S\to e^{i\alpha}S.
}
All these matrices induce the same automophism $\tau$ on the group $G_H$.

Now suppose we have at our disposal two GCP transformations \eqref{phi:gcp}
defined by two unitary matrices $S_1$ and $S_2$. If we apply them in succession, we
obtain
\eq{
\phi(x)\to S_1\phi^*(\hx) \to S_1S_2^*\phi(x)\,.
}
This means that two different GCP transformations induce a (unitary) horizontal
transformation 
\eq{
\phi(x)\to U\phi(x)\,,
}
with $U=S_1S_2^*$.

We distinguish two cases: (i) If this new horizontal $U$ transformation corresponds
to a representation $D(g)$ of an element $g$ in $G_H$, then only one of the two GCP
transformations $\tS_1,\tS_2$ has to be included as an additional transformation
subjected to the constraints \eqref{D_S=D:tau} and \eqref{SS^*}.
(ii) If the horizontal transformation $U$ does not correspond to an element of
$G_H$, then such a group has to be extended to a larger group $G_H'$. The simplest
way to extend $G_H$ to $G_H'$ is by split extension of the form
$G_H\rtimes\aver{U}$. In this case, $U$ also induces an automorphism $\tau$, by a 
unitary version of the transformation \eqref{D_S=D:tau}, as
\eq{
\label{D:U}
D_U(g)\equiv UD(g)U^{-1}=D(\tau(g))\,.
}
We denote this unitary transformation by $D_U$ without the tilde symbol.
We can also understand this requirement by the successive application of
$\tS_1,\tS_2$ as
\eq{
D(g)\to D_{\tS_1}(g)=S_1D(g)^*S_1^{-1}\to
D_{\tS_2\tS_1}(g)=S_2D_{\tS_1}(g)^*S_2^{-1}=S_2S_1^*D(g)S_1^\tp S_2^\dag\,.
}
We identify $U=S_2S_1^*$ and $\tau=\tau_2\circ\tau_1$ in \eqref{D:U} if $\tau_k$
is induced by $\tS_k$, $k=1,2$.
In these cases, the action of the antiunitary transformations $\tS_1$ and $\tS_2$,
inducing automorphisms $\tau_1,\tau_2$, is equivalent to the action of the unitary
transformation $U$ which induces the combined automorphism $\tau_2\circ\tau_1$.
We can equally compose unitary automorphisms with antiunitary automorphisms as well
as unitary ones with another unitary transformation.
The set of all matrices $S$ in \eqref{D_S} and $U$ in \eqref{D:U} represents
the automorphism group $\aut(G_H)$.

\section{The case of $G_H=\Delta(27)$}
\label{sec:D27}

The automorphism group of $\Delta(27)$ was discussed in Ref.\,\cite{lindner:gcp}.
The structure of the group is
\eqali{
\aut(\Delta(27))&\simeq
  \big(\big(\big(\ZZ_3{\times}\ZZ_3\big){\rtimes}Q_8\big)
  {\rtimes}\ZZ_3\big){\rtimes}\ZZ_2
~,&&
Z\equiv\mathcal{Z}\big(\Delta(27)\big)\simeq\ZZ_3,\cr
  \mathrm{Inn}(\Delta(27))&\simeq\Delta(27)/\mathcal{Z}\big(\Delta(27)\big)
  \simeq\ZZ_3\times\ZZ_3
~,&&
\out(\Delta(27))\simeq GL_2(\mathbb{F}_3)
  \simeq\big(Q_8{\rtimes}\ZZ_3\big){\rtimes}\ZZ_2\,.
}
Recall that the outer
automorphism group is defined by \eqref{outer}.
The possible nontrivial GCP transformations have to be associated with one of
the 48 elements of $\out(\Delta(27))$.

We can study $\aut(\Delta(27))$ by using the explicit representation
\eqref{D27:3} for the generators $a,b,a'$ in \eqref{D27:present}.
But instead of using $a'$, we can use \eqref{z0} as an auxiliary generator.
For the representation \eqref{D27:3}, we have
\eq{
\label{z:0}
z_0\sim \om^2\id_3\,.
} 
We can replace the presentation \eqref{D27:present} with
\eq{
\label{D27:present:2}
\begin{gathered}
a^3=b^3=z_0^3=e\,,~~az_0=z_0a\,,~~\cr
bab^{-1}=az_0\,,~~bz_0b^{-1}=z_0\,.
\end{gathered}
}

We can write all 27 elements of $\Delta(27)$ as
\eq{
g=b^{n_1}a^{n_2}z_0^{n_3}\,,
}
where $n_1,n_2,n_3$ runs from 0 to 2.

\subsection{Auxiliary result}
\label{sec:aux}

Let us show that for $G=\Delta(27)$, the following is true
\eq{
\label{aux}
\out(G)\equiv\aut(G)/\mathrm{Inn}(G)\simeq\aut(G/\mathcal{Z}(G))\,.
}
This means that to study the outer automorphism group of $\Delta(27)$, all we need
to know is the automorphism group of the smaller group
$\Delta(27)/Z\simeq\ZZ_3\times\ZZ_3$.
This property is very particular to $\Delta(27)$ and it is not satisfied, for
example, for $G=A_4$ or a cyclic group.

To establish \eqref{aux}, it is useful to define a homomorphism from $\aut(G)$
to $\aut(G/Z)$ by mapping $\tau\in\aut(G)$ to $\tau'\in\aut(G/Z)$ by
\eq{
\label{tau->tau'}
\tau'(xZ)=\tau(x)Z\,,
}
where $xZ\in G/Z$ is a coset of $Z$ in $G$, containing $x$ in $G$.
By the homomorphism theorem of group theory, we prove \eqref{aux} by showing 
that the kernel of this homomorphism is $\mathrm{Inn}(G)$.

By definition, the kernel of the homomorphism defined by \eqref{tau->tau'} is given
by automorphisms $\tau$ of $G$ mapped to $\tau'=\mathrm{id}$. This means
\eq{
\tau'(xZ)=\tau(x)Z=xZ, \text{ for all $xZ$ in $G/Z$.}
}
Then $\tau(x)z'=xz''$ for some $z',z''$ in $Z$. Finally, $\tau$ in the kernel of
the homomorphism \eqref{tau->tau'} is required to obey
\eq{
\label{tau:h}
\tau(x)=xz, \text{ for some $z$ in $Z$ and for all $x$ in $G$}.
}

The remaining task is to show that any automorphism $\tau$ of $G$ that obeys
\eqref{tau:h} is an inner automorphism. We do it explicitly for $G=\Delta(27)$ by
considering its generators $a,b$.
Any automorphism $\tau$ that obeys \eqref{tau:h} should be entirely determined by
how it acts on the generators $a,b$, i.e.,
\eq{
\label{tau:ab}
\tau(a)=az_0^n\,,~~
\tau(b)=bz_0^m\,.
}

Let us now show that any automorphism of the type \eqref{tau:ab} is an inner
automorphism. We begin by confirming that the validity of the property
$bab^{-1}=az_0$ in \eqref{D27:present:2} through the automorphism $\tau$ implies
$\tau(z_0)=z_0$. Then we check the action through conjugation of
\eq{
\label{conj:ab}
c_a(b)=bz_0^2\,,~~c_b(a)=az_0\,,
}
where we use the notation $c_g(x)=gxg^{-1}$ for conjugation.
Conjugation by $a,b$ in \eqref{conj:ab} allows us to compute the conjugation by a
general $g=b^{n_1}a^{n_2}z_0^{n_3}$ as
\eq{
c_g(a)=az_0^{n_1}\,,~~c_g(b)=bz_0^{-n_2}\,.
}
We made use of the property $b^na^m=a^mb^nz_0^{mn}$.
We can then conclude that any automorphism \eqref{tau:ab} corresponds to a
conjugation by some $g$ in $G$, and conversely any inner automorphism will have the
form \eqref{tau:ab}. This result establishes \eqref{aux} for $\Delta(27)$.

\subsection{Automorphism group of $\Delta(27)/\ZZ_3$}
\label{sec:Z3Z3}

Let us study the automorphism group of $\Delta(27)/Z\simeq \ZZ_3\times\ZZ_3$, where
$Z\equiv\mathcal{Z}(\Delta(27))\simeq\ZZ_3$ is the center of $\Delta(27)$. The
group $\Delta(27)/Z$ is generated by the cosets $\bar{a}=aZ$ and $\bar{b}=bZ$ while
$Z$ is generated by $z_0$ in \eqref{z0}.
An automorphism $\tau$ in $\Delta(27)/Z$ can be defined by knowing the mapping of
the generators $(\bar{a},\bar{b})\mapsto(\tau(\bar{a}),\tau(\bar{b}))$.
Part of this discussion can be also found in Ref.\,\cite{ivanov:3hdm}.

Next we know $\aut(\ZZ_3\times\ZZ_3)\simeq GL_2(\FF_3)$, the group of $2\times 2$
invertible matrices with entries in the finite field $\FF_3=\{-1,1,0\}$. We identify
$\Delta(27)/Z$ and $\ZZ_3\times\ZZ_3$ as follows: for each element
$\bar{x}=\bar{a}^{n}\bar{b}^{m}$ in $\Delta(27)/Z$ we define a vector in
$\FF_3^2=\FF_3\times\FF_3$ as
\eq{
\label{Z3->F3}
\bar{x}=\bar{a}^{n}\bar{b}^{m}\to \bp=(n,m)^\tp\,,
}
where $n,m=-1,0,1$. 
For example,
\eq{
\label{ab:F3}
\bar{a}\to (1,0)\,,~~\bar{b}\to (0,1)
\text{ and } \bar{a}\bar{b}^2\to (1,-1)=(1,0)+(0,-1)\,.
}
Therefore, we trade group multiplication in $\ZZ_3\times\ZZ_3=
\aver{\bar{a}}\times\aver{\bar{b}}$ for vector addition in $\FF_3^2$.
Now a matrix $A$ in $GL_2(\FF_3)$ induces an autormophism in
$\FF_3^2$ by
\eq{
\bp\to A\bp\,.
}
The automorphism on $\Delta(27)/Z$ can be read off from \eqref{Z3->F3}.
For example,
\eq{
\label{A:tau}
\tau\sim A=\mtrx{1&1\cr-1&0} \text{ is equivalent to }
\left\{
\begin{array}{rcl}
\tau(\bar{a})&=&\bar{a}\bar{b}^2\cr
\tau(\bar{b})&=&\bar{a}\,.
\end{array}
\right.
}

We can split $GL_2(\FF_3)$ into $SL_2(\FF_3)\rtimes\ZZ_2$ where $\ZZ_2$ is generated
by a $2\times 2$ matrix of determinant $-1$, associated to the automorphism
$\sigma$.
Let us choose
\eq{
\label{def:d}
\sigma\sim d\equiv\mtrx{-1&0\cr0&1}\,,
}
so that $\aut(\Delta(27)/Z)\simeq SL_2(\FF_3)\rtimes\aver{\sigma}$.
Now we only need to study the subgroup isomorphic to $SL_2(\FF_3)$.

Let us show that $SL_2(\FF_3)\simeq Q_8\rtimes\ZZ_3$ by picking up some
elements of $SL_2(\FF_3)$,
\eq{
\label{def:ei:c}
\begin{gathered}
e_1\equiv\mtrx{1&1\cr1&-1}\,,~
e_2\equiv\mtrx{-1&1\cr1&1}\,,\cr
e_3\equiv\mtrx{0&-1\cr1&0}\,,~
c\equiv\mtrx{1&1\cr0&1}\,.
\end{gathered}
}
We can show that 
\eq{
\label{Q8:Z3}
\aver{e_1,e_2,e_3}\simeq Q_8\,,~~ \aver{c}\simeq\ZZ_3\,,
}
and that $\aver{e_1,e_2,e_2}\aver{c}$ generates an order 24 group which
exhausts $SL_2(\FF_3)$.

Firstly, we can directly show that the following properties hold:
\eq{
\begin{gathered}
e_1^2=e_2^2=e_3^2=-\id_2\,,~~\cr
e_1e_2=-e_2e_1=e_3\,,~~c^3=\id_2\,.
\end{gathered}
}
These properties establish \eqref{Q8:Z3}.
The semidirect product $Q_8\rtimes\ZZ_3$ is confirmed from the automorphism on
$Q_8$ generated by $c$ as 
\eq{
ce_1c^{-1}=e_2\,,~~
ce_2c^{-1}=e_3\,,~~
ce_3c^{-1}=e_1\,.
}
Finally, we can check that $\aver{e_1,e_2,e_2}$ and $\aver{c}$ have trivial
intersection and that $\aver{e_1,e_2,e_2}\aver{c}$ has 24 elements.

For completeness, we can add the element $d$ in \eqref{def:d} to generate
$GL_2(\FF_3)$. The element $d$ induces an automorphism on $SL_2(\FF_3)$ as
\eq{
\begin{gathered}
de_1d^{-1}=e_2^{-1}\,,~~
de_2d^{-1}=e_1^{-1}\,,\cr
de_3d^{-1}=e_3^{-1}\,,~~
dcd^{-1}=c^{-1}\,.
\end{gathered}
}
We can write all elements of $GL_2(\FF_3)$ as a product of an element in
$SL_2(\FF_3)$ and $\id_2$ or $d$.

\subsection{Unitary and antiunitary automorphisms}

We show here the following result: antiunitary transformations \eqref{D_S}
induce automorphisms in $\out(\Delta(27))$ corresponding to matrices $A$
in $GL_2(\FF_3)$, with $\det A=-1$, and unitary transformations \eqref{D:U} realize 
automorphisms corresponding to elements of $SL_2(\FF_3)$.

Given that the center of a group is always mapped into itself by any automorphism,
we can firstly distinguish two types of automorphisms in $\aut(\Delta(27))$:
\eq{
\text{(I) }  \tau(z_0)=z_0\,,\qquad
\text{(II) }  \tau(z_0)=z_0^{-1}\,.
}
We can easily see that automorphisms of type I form a normal subgroup
of $\aut(\Delta(27))$ with half of the elements.
In general, we would call a CP-type transformation a type II transformation which
sends $z_0\to z_0^{-1}$. We can see that by noting that only the triplet and
antitriplet representations, $\bs{3}$ and $\bbs{3}$, represent the element $z_0$ of
the center nontrivially. For example, for $\bs{3}$, with the choice \eqref{D27:3} we
obtain \eqref{z:0} for $z_0$. Since $z_0$ is in the center, its representation
$\bs{3}$ is proportional to the identity and we can immediately see that the
automorphisms induced by a unitary transformation \eqref{D:U} are of type I whereas
the automorphisms induced by \eqref{D_S} are of type II.

Our task is to show that the subgroup of type I automorphisms coincide with the
subgroup $SL_2(\FF_3)$ of $\aut(\Delta(27))$ modulo inner automorphisms. We follow
Ref.\,\cite{ivanov:3hdm}, Sec.\,7.1.
The first step is to define the commutator of two elements $x,y$ of the group
\eq{
[x,y]\equiv xyx^{-1}y^{-1}\,.
}
This operation has the properties
\eq{
\label{[]:prop}
[y,x]=\big([x,y]\big)^{-1}~ \text{ and }~
[xx',y]=[x,y][x',y]\,,
}
where the last relation is already specialized to $\Delta(27)$ where all commutators
lie in the center $Z$.
For example,
\eq{
[b,a]=bab^{-1}a^{-1}=z_0\,.
}
We can also identify the commutator in $\Delta(27)$ and $\Delta(27)/Z$ as
\eq{
[x,y]=[\bar{x},\bar{y}]\,,
}
since the commutator is invariant if we replace $x$ with $xz$, where $z\in Z$.
The same is true for $y$.

The next step is to use the mapping \eqref{Z3->F3} to define a bilinear
$d$ function of $\FF_3^2$ to $\FF_3$ by
\eq{
d(\bp,\bq)=n,~\text{ from }~[\bar{x},\bar{y}]=z_0^n\,,~
\text{ if } \bp\to\bar{x},\,\bq\to\bar{y}\,.
}
The integer $n=-1,0,1$ belongs to $\FF_3$.
The properties \eqref{[]:prop} translate to the following properties of $d$:
\eq{
d(\bq,\bp)=-d(\bp,\bq),~~
d(\bp+\bp',\bq)=d(\bp,\bq)+d(\bp',\bq)\,.
}
Thus $d$ is a bilinear function.
Since $[\bar{a},\bar{b}]=z_0^2\sim d((1,0),(0,1))=-1$, $d(\bp,\bq)$ corresponds
to $(-1)$ times the determinant of the matrix formed by columns
$\bp,\bq$ (from the unicity of the determinant function). This observation leads to
\eq{
d(A\bp,A\bq)=\det(A)d(\bp,\bq)\,,
}
where $A\in GL_2(\FF_3)$.
Finally, we can see how an automorphism $\tau$ associated to a matrix $A$ acts on
$z_0=[\bar{b},\bar{a}]\sim d((0,1),(1,0))=1$:
\eq{
z_0\to d(A(0,1),A(1,0))=\det(A)\sim 
[\tau(\bar{b}),\tau(\bar{a})]=\tau([\bar{b},\bar{a}])=\tau(z_0)
=(z_0)^{\det(A)}\,.
}
Hence elements of $SL_2(\FF_3)$ [$GL_2(\FF_3)-SL_2(\FF_3)$] act as type I
[type II] automorphisms.

\subsection{Obtaining the matrices $S_i$}
\label{sec:obtaining}

We are now in the position to calculate the matrices $S$ that induce the
automorphisms in \eqref{D_S=D:tau} for the triplet representation $\bs{3}$,
Eq.\,\eqref{D27:3}. These matrices will define the GCP transformations \eqref{gcp:L}
for the lepton doublets.

The relation \eqref{aux} allows us to associate, in a one-to-one fashion, a matrix
$A\in GL_2(\FF_3)$ to each automorphism $\tau$ of $\aut(\Delta(27))$, modulo inner
automorphisms. For GCP transformations, we only need the elements of
$GL_2(\FF_3)-SL_2(\FF_3)$, with determinant $(-1)$, which can be written as
\eq{
\label{A.d}
A=A'd\,,
}
where $A'\in SL_2(\FF_3)$ and $d$ was defined in \eqref{def:d}; see
Sec.\,\ref{sec:Z3Z3}.
In turn, all the elements $A'$ of $SL_2(\FF_3)$ can be recovered from the structure
$\aver{e_1,e_2,e_3}\rtimes\aver{c}\simeq Q_8\rtimes\ZZ_3$ whose generators were
defined in Eq.\,\eqref{def:ei:c}.

We can calculate all the possible matrices $S$ that defines GCP
transformations by using \eqref{D_S=D:tau} for elements of $GL_2(\FF_3)$
of the type \eqref{A.d}.
Let us begin by the simplest $A=d$ case. We need a matrix $S(d)$ that
induces automorphism $(a,b)\to (a^2,b)$, i.e.,
\eq{
\label{S:d}
S(d)D_3(a)^*S(d)^{\dag}=D_3(a^2)\,,~~
S(d)D_3(b)^*S(d)^{\dag}=D_3(b)\,.
}
We use the triplet representation \eqref{D27:3}.
Since \eqref{S:d} requires that $S(d)$ commutes with both $D_3(a)$ and $D_3(b)$, 
the only solution is
\eq{
S(d)=\id_3\,,
}
neglecting a possible phase factor. Thus the GCP associated to the automorphism $d$
is just the usual CP transformation.

We can obtain all other matrices $S$ by composition from \eqref{A.d} since
\eq{
\label{D:A'd}
D((A'd)(g))=U(A')D(d(g))U(A')^\dag=U(A')D(g)^*U(A')^\dag
\,,
}
where we have denoted as $(A'd)(g)$ the element mapped by automorphism from
$g$ by the matrix $A'd$, using some convention explained below. For the generators
$a,b$ we need
\eq{
\label{S:A}
U(A')D_3(a)U(A')^{\dag}=D_3(A'(a))\,,~~
U(A')D_3(b)U(A')^{\dag}=D_3(A'(b))\,.
}
We seek only matrices that can be associated to one single GCP transformation, which
requires \eqref{SS^*}, \eqref{tau:2:2} and \eqref{tau:s}.
In special, \eqref{tau:2:2} implies we only need to consider order two automorphisms
in $\out(\Delta(27))$.

All the order two automorphisms of $\out(\Delta(27))$ are in the conjugacy class of
$d$. Such a conjugacy class is composed by the 12 elements
\eq{
\label{C(d)}
\bigg\{\mtrx{-1&0\cr0&1},\mtrx{-1&1\cr0&1},\mtrx{-1&-1\cr0&1},
\mtrx{-1&0\cr1&1},\mtrx{-1&0\cr-1&1},\mtrx{0&1\cr1&0}\bigg\}
\Big\{\id,-\id_2\Big\}\,.
}
We can denote the elements within the first braces by
$\{d,cd,c^2d,-e_2cd,e_1c^2d,-e_3d\}$, respectively.

To find all the matrices $S$ corresponding to the automorphisms \eqref{C(d)}, we
only need to find the unitary matrices $U$ satisfying \eqref{S:A} for the
automorphisms $\{\id_2,c,c^2,-e_2c,e_1c^2,-e_3\}\{\id_2,-\id_2\}$, which
corresponds to the set \eqref{C(d)} multiplied by $d$ from the right.
To construct all of them, we only need $\{-\id_2,c,-e_2c,e_3\}$, i.e.,
\eq{
\bigg\{\mtrx{-1&0\cr0&-1},\mtrx{1&1\cr0&1},\mtrx{1&0\cr-1&1},\mtrx{0&-1\cr1&0}
  \bigg\}
\,,
}
respectively; the rest can be obtained from their inverses.

Note that automorphism $A'$ is unique in $\out(\Delta(27))$. In
$\aut(\Delta(27))$, they are defined up to inner automorphisms.
To define an automorphism in $\aut(\Delta(27))$ from $GL_2(\FF_3)$ we need a
convention, i.e., a recipe to extract one representative element from the coset.
We adopt the following: we drop the bar in \eqref{A:tau} and define the mapping
on $a,b$. 
One only needs to define an ordering for terms with products of $a$ and $b$.
We use the ordering $b^{n_2}a^{n_1}$.
For example, for the automorphism $c$ above, we seek a $U(c)$ that induces
\eq{
\label{conv}
c=\mtrx{1&1\cr0&1}:~~(a,b)\mapsto (a,ba).
}
Thus, we conveniently write $c(a)=a,~c(b)=ba$ as in \eqref{D:A'd}.

Imposing \eqref{S:A}, we find
\eqali{
\label{U:A}
U(-\id_2)&=\mtrx{1&0&0\cr0&0&1\cr0&1&0}\,,&
  U(e_3)&=\frac{1}{\sqrt{3}}\mtrx{1&1&1\cr1&\om&\om^2\cr1&\om^2&\om}
    \equiv U_\om\,,\cr
U(c)&=\mtrx{1&0&0\cr0&1&0\cr0&0&\om}\,,&
  U(-e_2c)&=\frac{-i\om}{\sqrt{3}}\mtrx{1&\om^2&1\cr1&1&\om^2\cr\om^2&1&1}
    \equiv U_3'\,.
}
Note that for $-e_2c$ we have used $e_3ce_3^{-1}=-e_2c$, so that the convention
\eqref{conv} is not respected.
To obtain a symmetric matrix, we redefine
\eq{
U(-e_2c)=D_3(b)U_3'=
  \frac{-i\om}{\sqrt{3}}\mtrx{\om^2&1&1\cr1&\om^2&1\cr1&1&\om^2}
  \equiv U_3\,.
}

Finally, the list of matrices \eqref{list:S} is obtained from \eqref{U:A} as
follows:
\eq{
\begin{gathered}
S_0=\id_3,~~S_1=U(-\id_2),~~S_2=U(c),~~S_3=U(c)^{-1},\cr
S_4=U(e_3),~~S_5=U(e_3)S_1,~~S_6=U_3S_1, ~~S_7=U_3^{-1}S_1,\cr
S_8=TS_2T^{-1}S_1,~~S_9=TS_3T^{-1}S_1,~~S_{10}=U_3, ~~S_{11}=U_3^{-1}.
\end{gathered}
}
To define symmetric $S_8,S_9$, we have conveniently included the inner
automorphism $T=D_3(b)$. 
All GCP transformations in $\Delta(27)$ models are then defined by the matrices
$S_i$, $i=0,\ldots,11$, up to inner automorphisms.
The choice of symmetric $S_i$ implies that the associated automorphisms $\tau_i$
have order two in $\aut(\Delta(27))$, as in \eqref{s=e}, and not only in
$\out(\Delta(27))$.
Therefore, in $\Delta(27)$, we can define all GCP transformations in terms of
symmetric $S_i$, which are the symmetries relevant for residual GCP symmetries on
the mass matrix $M_\nu$.
However, this does not mean that we can not define GCP transformation with
nonsymmetric $S$. For example, defining $S_{10}=U_3'$ in \eqref{U:A} instead of
$U_3$ leads to
\eq{
S_{10}S_{10}^{*}=\mtrx{0&0&1\cr1&0&0\cr0&1&0}=D_3(b)\,.
}
In this case, \eqref{SS^*} is valid with nontrivial $s=b$.

We should also emphasize that the matrices in \eqref{U:A} are defined up to phases.
If we are only interested in GCP transformations, Eq.\,\eqref{S:e^i} ensures that
phases are unimportant. Instead, if we want to enlarge the $\Delta(27)$ group
(identifying it with its triplet representation) by the inclusion of some of
the elements above, then the phase factors should be compatible with the order of
the element.
For example, $U(c)$ is defined up to factors $1,\om,\om^2$ while $U(e_3)$ can be
multiplied only by $\pm1,\pm i$.
For completeness, the order of $-e_2c$ is also 3.

Let us finish the study of GCP in $\Delta(27)$ flavor models by giving an
explicit example showing that, for $\Delta(27)$, the consistency condition
\eqref{SS^*} is \textit{additional} to the automorphism condition
\eqref{D_S=D:tau}. 
If we take the automorphism associated with $\tau(a)=ba,\tau(b)=a$, we have
$\tau^8=\mathrm{id}$ modulo conjugation.
If we solve the condition \eqref{D_S=D:tau} for $a,b$, we find
\eq{
\label{tau:4:example}
S=\frac{1}{\sqrt{3}}\mtrx{1&1&1\cr1&\om&\om^2\cr\om&1&\om^2}\,,
}
up to rephasing.
For the consistency condition \eqref{SS^*}, we find
\eq{
SS^*=\frac{i}{\sqrt{3}}\mtrx{\om^2&\om^2&1\cr1&\om^2&\om^2\cr1&\om&1}\,.
}
This element is not part of $\Delta(27)$, so the horizontal group needs to
be enlarged by including it.
One can check that $(SS^*)^4=\id_3$.

\section{Conclusions}
\label{sec:conclusion}

We have found all GCP transformations that can be defined in $\Delta(27)$
flavor models, up to composition with elements of $\Delta(27)$ or multiplication by
a phase factor. The list is shown in \eqref{list:S}. The inclusion of any other GCP
transformation leads to the enlargement of the flavor group $\Delta(27)$.
Moreover, the extension of the flavor group by any GCP transformation
is equivalent to the addition of an antiunitary transformation \eqref{gcp:L} for
which the unitary part $S$ that connects different families is symmetric.

We have also discussed the consequences of having a \textit{single} GCP symmetry as
a residual symmetry of the neutrino mass matrix $M_\nu$. In the flavor basis, the
presence of a CP-type residual symmetry of $M_\nu$ fixes three out of the 6
parameters of the leptonic mixing matrix, more precisely, three complex phases
that lead indirectly to the Dirac CP phase and the two Majorana phases.
Although the mixing angles are unconstrained, there is an intrinsic part of the
mixing matrix that is completely determined by the GCP residual symmetry. If other
symmetries are able to ensure that the unconstrained part is near the identity
matrix, then the residual symmetry is capable of fixing the approximate features of
the leptonic mixing matrix.

Specifically for $\Delta(27)$ flavor models, we have identified some potential GCP
symmetries that lead to interesting patterns for the parts that are determined by
the residual symmetry. In particular, two GCP symmetries lead to the tribimaximal
form for the intrinsic part of the PMNS matrix. Another two GCP symmetries lead to a
new pattern numerically close to but distinct from the tribimaximal form.
These patterns could be further employed in flavor model building to explain the
observed mixing patterns of leptons, including the nonzero $\theta_{13}$ angle.

\acknowledgements

The work of the author was partially supported by Brazilian CNPq and Fapesp.
The author also thanks R.~N.~Mohapatra for fruitful discussions.


\end{document}